\documentstyle[12pt]{article}
\begin{document}
\title{NEUTRINO MASS AND AN EVER EXPANDING UNIVERSE\\
(AN IRREVERENT PERSPECTIVE)$^+$}
\author{B.G. Sidharth$^*$\\
Centre for Applicable Mathematics \& Computer Sciences\\
B.M. Birla Science Centre, Adarsh Nagar, Hyderabad - 500063 (India)}
\date{}
\maketitle
\footnotetext{\noindent E-mail:birlasc@hd1.vsnl.net.in\\
$^+$Invited talk at Workshop on Neutrino Physics, University of Hyderabad,
November, 1998}
\begin{abstract}
There have been two significant recent findings. One is a culmination of the
Superkamiokande experiments, which demonstrate Neutrino oscillation and therefore
a non-zero mass. The other is the finding that the universe will continue to
expand without declaration based on distant supernovae observations. At the
very least these two findings call for a review of the existing and generally
accepted theories. In this talk it is pointed out how such a Neutrino Mass
can in fact be deduced from a theoretical model, as also the eternal expansion feature
of the universe, and how both these findings do not contradict each other.
\end{abstract}
\section{Introduction and Review}
We start with a brief background on the neutrino\cite{r1}-\cite{r7}.
The neutrino was proposed in 1929 by W. Pauli. In his words,
"Dear Radioactive Ladies and Gentlemen,..... as a desperate
remedy to save the principle of energy conservation in $\beta$
decay, ..... I propose the idea of a neutral particle of spin
half."\\
A few years later Fermi introduced the four Fermi Hamiltonian for $\beta$
decay using the Neutrino and so the theory of weak interactions
was born.\\
Finally in 1956 Reines and Cowan discovered the Neutrino.\\
The Neutrino turned out to be a massless, chargeless spin half
particle with handedness. There are an estimated $10^{90}$ neutrinos
in the universe.\\
From the beginning it has been an enigmatic particle. In the words
of Pauli, again, nearly thirty years after he had first postulated it
"this particle neutrino, of the existence of which
I am not innocent, still persecutes me."\\
The generally accepted standard model retains these features of
the neutrino, except that there are three families, the electron,
the muon and the tau neutrinos. Because of their vanishing mass,
mixings and magnetic moment also vanish. However higher order
weak interactions endow it with a  charge radius. Infact the
electromagnetic processes appear only in the elastic scattering of
neutrinos with electrons or quarks.\\
Though the standard model predicts zero neutrino mass, all that
we can say from experiment is that there are the following upper
bounds for the neutrino mass:
$$\mbox{Mass of}\nu_e < 12eV$$
$$\mbox{Mass of}\nu_\mu < 170keV$$
$$\mbox{Mass of} \nu_\tau < 24MeV$$
This could provide a motivation for a study of physics beyond a
standard model. Further there is no apriori theoretical reason
why the right handed neutrino field should not exist, unlike in the
case of the masslessness of the photon. In other words there are
only left handed neutrinos in the standard model, just to conform
to observation.\\
On the other hand, many unification schemes do predict a neutrino
mass. Though the idea of a neutrino mass and neutrino oscillation
goes a long way back to authors like Markov, Pontecarvo and others,
several puzzling questions have persisted. These include, the question
of the smallness of the neutrino mass and the fact that the observed
number of solar neutrinos is less than half the expected number. This
latter problem would be solved if the neutrinos are a superposition
of different mass Eigen states, leading to neutrino oscillation
and therefore mass.\\
As suitable mass for the neutrino would also solve the problem of
dark matter in the galaxies, and it could also resolve the problem
of the missing mass of the universe. This is because standard big
bang cosmology predicts the existence of relic background neutrinos.
If these neutrinos had a mass of about $10eV$, the universe
would be closed.\\
Though neutrinos are considered not to have any electric charge,
they could have a magnetic moment induced by quantum loops in electro
weak theories. Infact a magnetic moment $\sim 10^{-10} \mu_B$
would also solve the solar neutrino puzzle. This is because the
neutrinos can undergo a spin flip in the sun's magnetic field, and
go over to righthanded neutrinos, which because of their extremely
weak interactions cannot be detected. However, this value appears
to be very high and theory predicts a value $<_{\sim} 10^{-19}\mu_B$.\\
Interestingly the Kamiokande experiments based on observations
of the supernova SN 1987 A indicate that if the neutrino has a
charge at all, this will be less than or equal to $10^{-17}e$,
rather than the earlier limit, greater than or equal to $10^{-13}e$.\\
It may also be mentioned that recent versions of the neutrino
as a Dirac particle suggest a mass given by,
$$m_\nu \approx 10^{-7} m_l,$$
where
$m_l$ is the relevant lepton.\\
Finally we sum up the experimental evidence which suggested a
non zero neutrino mass before the recent SuperKamiokande experiments:\\
i) Solar neutrino deficit.\\
ii)The deficit of muon neutrinos relative to electron neutrinos
produced in the atoms.\\
iii) The neutrino oscillation observed at the Los Alamos Liquid
Scintillation Neutrino Detector (LSND).\\
iv) The Russian Tritium experiment.\\
v) The astronomical input in the form of dark matter or missing mass.\\
The latest experiments suggest a neutrino mass $\sim 10^{-8}$
electron mass.
\section{Kerr-Newman Formulation and Consequences}
We now consider the neutrino in a slightly different context.
According to a recent model, elementary particles, typically leptons,
can be treated as, what may be called Quantum Mechanical Black Holes
(QMBH)\cite{r8}-\cite{r12}, which share certain features of Black Holes and also
certain Quantum Mechanical characteristics. Essentially they are bounded
by the Compton wavelength within which non local or negative energy phenomena
occur, these manifesting themselves as the Zitterbewegung of the electron.
These Quantum Mechanical Black Holes are created out of the background Zero
Point Field and this leads to a consistent cosmology, wherein using $N$, the
numbef of particles in the universe as the only large scale parameter, one
could deduce from the theory, Hubble's law, the Hubble constant,
the radius, mass, and age of the universe and features like the hitherto
inexplicable relation between the pion mass and the Hubble constant
\cite{r8}. The model also predicts an ever expanding universe, as
recent observations do confirm.\\
Within this framework, it was pointed out that the neutrino would be a mass
less and charge less version of the electron and it was deduced that it would
be lefthanded, because one would everywhere encounter the pseudo spinorial
("negative energy") components of the Dirac spinor, by virtue of the fact
that its Compton wavelength is infinite (in practise very large). Based on
these considerations we will now argue that the neutrino would exhibit an
anomalous Bosonic behaviour which could provide a clue to the
neutrino mass.
\section{The anomalous neutrino}
As detailed in Ref.\cite{r9} the Fermionic behaviour is due to the non local
or Zitterbewegung effects within the Compton wavelength effectively showing
up as the well known negative energy components of the Dirac spinor
which dominate within while positive energy components predominate
outside leading to a doubly connected space or equivalently the spinorial or
Fermionic behaviour.
In the absence of the Compton wavelength boundary, that
is when we encounter only positive energy or only negative energy solutions,
the particle would not exhibit the double valued spinorial or Fermionic behaviour:
It would have an anomalous anyonic behaviour.\\
Indeed, the three
dimensionality of space arises from the spinorial behaviour outside the
Compton wavelength\cite{r13}. At the Compton wavelength, this disappears
and we should encounter lower dimensions. As is well known\cite{r14} the
low dimensional Dirac equation has like the neutrino, only two
components corresponding to only one sign of the energy, displays
handedness and has no invariant mass.\\
Ofcourse the above model strictly speaking is for the case of an isolated
non interacting particle. As neutrinos interact through the weak or
gravitational forces, both of which are weak, the conclusion would still be
approximately valid particularly for neutrinos which are not in bound states.\\
We will now justify the above conclusion from three other standpoints.:
Let us first examine why Fermi-Dirac statistics is required in the Quantum
Field Theoretic treatment of a Fermion satisfying the Dirac equation. The
Dirac spinor has four components and there are four independent solutions
corresponding to positive and negative energies and spin up and down. It
is well known that\cite{r15} in general the wave function expansion of the
Fermion should include solutions of both signs of energy:

\begin{eqnarray}
\psi (\vec x,t) & = & N \int d^3 p \sum_{\pm s} [b(p,s)u(p,s)\exp (-\imath p^\mu
x_\mu/\hbar)\nonumber\\
& & + d^*(p,s)v(p,s) \exp (+\imath p^\mu x_\mu/\hbar)\label{e1}
\end{eqnarray}
where $N$ is a normalization constant for ensuring unit probability.\\
In Quantum Field Theory, the coefficients become creation and annihilation
operators while $bb^+$ and $dd^+$ become the particle number operators with
eigen values $1$ or $0$ only. The Hamiltonian is now given by\cite{r16}:
\begin{equation}
H = \sum_{\pm s} \int d^3 pE_p[b^+(p,s)b(p,s) - d(p,s)d^+(p,s)]\label{e2}
\end{equation}
As can be seen from (\ref{e2}), the Hamiltonian is not positive definite and
it is this circumstance which necessitates the Fermi-Dirac statistics. In
the absence of Fermi-Dirac statistics, the negative energy states are not
saturated in the Hole Theory sense so that the ground state would have
arbitrarily large negative energy, which is unacceptable. However Fermi-Dirac
statistics and the anti commutators implied by it prevent this from
happening.\\
From the above, it follows that as only one sign of energy is
encountered for the $\nu$, we need not take recourse to Fermi-Dirac
statistics.\\
We will now show from an alternative view point also that for the neutrino, the positve and negative solutions
are delinked so that we do not need the negative solutions in (\ref{e1})
or (\ref{e2}) and there is no need to invoke Fermi-Dirac statistics.\\
The neutrino is described by the two component Weyl equation\cite{r17}:
\begin{equation}
\imath \hbar \frac{\partial \psi}{\partial t} = \imath \hbar c\vec \sigma.
\vec \Delta \psi (x)\label{e3}
\end{equation}
It is well known that this is equivalent to a mass less Dirac particle
satisfying the following constraint (ref.\cite{r17}):
\begin{equation}
\Gamma_5 \psi = -\psi\label{e4}
\end{equation}
We now observe that in the case of a massive Dirac particle, if we work
only with positive solutions for example, the current or expectation value of the
velocity operator $c\vec \alpha$ is given by (ref.\cite{r15}),
\begin{equation}
J^+ = \langle c \alpha \rangle = \langle \frac{c^2 \vec p}{E}\rangle_+
= \langle v_{gp}\rangle_+\label{e5}
\end{equation}
in an obvious notation.\\
(\ref{e5}) leads to a contradiction: On the one hand the eigen values of
$c\vec \alpha$ are $\pm c$. On the other hand we require, $\langle v_{gp}\rangle
< 1$.\\
To put it simply, working only with positive solutions, the Dirac particle
should have the velocity $c$ and so zero mass. This contradiction is resolved by including the
negative solutions also in the description of the particle also. This infact is the starting point for (\ref{e1}) above.\\
In the case of mass less neutrinos however, there is no contradiction
because they do indeed move with the velocity of light. So we need not consider
the negative energy solutions and need work only with the positive solutions.\\
There is another way to see this. Firstly, as in the case of massive
Dirac particles, let us consider the packet (\ref{e1}) with both positive
and negative solutions for the neutrino. Taking the $z$ axis along the
$\vec p$ direction for simplicity, the acceptable positive and negative Dirac
spinors subject to condition (\ref{e4}) are
$$
u = \left(\begin{array}{c}
1\\0\\-1\\0\end{array}\right),\quad
v = \left(\begin{array}{c}
0\\-1\\0\\1\end{array}\right)
$$
The expression for the current is now given by,
\begin{eqnarray}
J^z = \int d^3 p \{\sum_{\pm s}[|b(p,s)|^2+|d(p,s)|^2]\frac{p^zc^2}{E}\nonumber\\
+ \imath \sum_{\pm s \pm s'} b^*(-p,s')d^*(p,s) - \bar u (-p,s')\sigma^{30}
\nu (p,s)\nonumber\\
- \imath \sum_{\pm s \pm s'} b(-p,s')d(p,s)-\bar \nu (p,s')\sigma^{30}u(-p,s)\}
\label{e6}
\end{eqnarray}
Using the expressions for $u$ and $v$ it can easily be seen that in
(\ref{e6}) the cross (or Zitterbewegung) term disappears.\\
Thus the positive and negative solutions stand delinked in contrast to the
case of massive particles, and we need work only with positive solutions (or
only with negative solutions) in (\ref{e1}).\\
Finally this can also be seen in yet another way. As is known (ref.\cite{r17}),
we can apply a Foldy-Wothuysen transformation to the mass less Dirac equation
to eliminate the "odd" operators which mix the components of the spinors
representing the positive and negative solutions.\\
The result is the Hamiltonian,
\begin{equation}
H' = \Gamma^o pc\label{e7}
\end{equation}
Infact in (\ref{e7}) the positive and negative solutions stand delinked. In
the case of massive particles however, we would have obtained instead,
\begin{equation}
H' = \Gamma^o \sqrt (p^2c^2 + m_oc^4)\label{e8}
\end{equation}
and as is well known, it is the square root operator on the right which gives
rise to the "odd" operators, the negative solutions and the Dirac spinors.
Infact this is the problem of linearizing the relativistic Hamiltonian and is
the starting point for the Dirac equation.\\
Thus in the case of mass less Dirac particles, we need work only with
solutions of one sign in (\ref{e1}) and (\ref{e2}). The equation (\ref{e2}) now becomes,
\begin{equation}
H = \sum_{\pm s} \int d^3 p E_p [b^+(p,s)b(p,s)]\label{e9}
\end{equation}
As can be seen from (\ref{e9}) there is no need to invoke Fermi-Dirac
statistics now. The occupation number $bb^+$ can now be arbitrary because the
question of a ground state with arbitrarily large energy of opposite sign does not arise.
That is, the neutrinos obey anomalous statistics.\\
In a rough way, this could have been anticipated. This is because the
Hamiltonian for a mass less particle, be it a Boson or a Fermion, is given by
$$H =\ pc$$
Substitution of the usual operators for $H$ and $p$ yields an equation in
which the wave function $\psi$ is a scalar corresponding to a Bosonic particle.
\section{The Spin Statistics Theory}
According to the spin-statistics connection, microscopic causality is
incompatible with quantization of Bosonic fields using anti-commutators and
Fermi fields using commutators(\cite{r16}). But it can be shown that this
does not apply when the mass of the Fermion vanishes.\\
In the case of Fermionic fields, the contradiction with microscopic causality
arises because the symmetric propogator, the Lorentz invariant function,
$$\Delta_1(x-x') \equiv \int \frac{d^3k}{(2\pi)^33\omega_k} [e^{-\imath k.(x-x')} +
e^{\imath k.(x-x')}]$$
does not vanish for space like intervals $(x-x')^2 < 0$, where the vacuum
expectation value of the commutator is given by the spectral representation,
$$S_1(x-x') \equiv \imath < 0|[\psi_\alpha (x), \psi_\beta (x')]|0 > = -
\int dM^2[\imath \rho_1(M^2)\Delta_x + \rho_2(M^2)]_{\alpha \beta} \Delta_1
(x-x')$$
Outside the light cone, $r > |t|$, where $r \equiv |\vec x - \vec x'|$ and
$t \equiv |x_o - x_o'|,\Delta_1$ is given by,
$$\Delta_1 (x'-x) = - \frac{1}{2\pi^2 r}\frac{\partial}{\partial r}
K_o(m\sqrt{r^2-t^2}),$$
where the modified Bessel function of the second kind, $K_o$ is given by,
$$K_o(mx) = \int^\infty_o \frac{\cos (xy)}{\sqrt{m^2+y^2}}dy = \frac{1}{2}
\int^\infty_{-\infty} \frac{\cos (xy)}{\sqrt{m^2+y^2}}dy$$
(cf.\cite{r18}). In
our case, $x \equiv \sqrt{r^2-t^2},$ and we have,
$$\Delta_1 (x-x') = const \frac{1}{x} \int^\infty_{-\infty}
\frac{y \sin xy}{\sqrt{m^2+y^2}}dy$$
As we are considering massless neutrinos, going to the limit as $m \to 0$, we
get, $|Lt_{m \to 0}\Delta_1 (x-x')| = |(const.). Lt_{m \to 0}\frac{1}{x}
\int^\infty_{-\infty} \sin xydy| < \frac{0(1)}{x}$. That is,
as the Compton wavelength for the neutrion is infinite (or very large), so
is $|x|$ and we have $|\Delta_1 | < < 1.$ So the
invariant $\Delta_1$ function nearly vanishes everywhere except on the light cone
$x = 0$, which is exactly what is required. So, the spin-statistics theorem
or microscopic causality is not violated for the mass less neutrinos when
commutators are used.\\
\section{Neutrino Mass}
The fact that the ideally, massless, spin half neutrino obeys anomalous
statistics could have interesting implications. For,
given an equilibrium  collection of neutrinos, we should have if we use
the Bose-Einstein statistics\cite{r19}.
\begin{equation}
PV = \frac{1}{3}U,\label{e10}
\end{equation}
instead of the usual
\begin{equation}
PV = \frac{2}{3}U,\label{e11}
\end{equation}
where $P,V$ and $U$ denote the pressure, volume and energy of the collection.
We also have, $PV \alpha NkT, N$ and $T$ denoting the number of particles
and temperature respectively.\\
On the other hand for a fixed temperature and number of neutrinos, comparison
of (\ref{e10}) and (\ref{e11})shows that the effective energy $U'$ of the
neutrinos would be twice the expected energy $U$. That is in effect the
neutrino acquires a rest mass $m$. It can easily be shown from the
above that,
\begin{equation}
\frac{mc^2}{k} <_{\approx} \sqrt{3}T\label{e12}
\end{equation}
That is for cold background neutrinos $m$ is about a thousandth
of an $ev$  at the present background temperature of about $2^\circ K$:\\
\begin{equation}
10^{-9} m_e \le m \le 10^{-8} m_e\label{e13}
\end{equation}
This can be confirmed, alternatively, as follows.
As pointed out by Hayakawa, the balance of the gravitational force and the Fermi energy of these
cold background neutrinos, gives\cite{r20},
\begin{equation}
\frac{GNm^2}{R} = \frac{N^{2/3}\hbar^2}{mR^2},\label{e14}
\end{equation}
where $N$ is the number of neutrinos.\\
Further as in the Kerr-Newman Black Hole formulation equating
(\ref{e14}) with the energy of the neutrino, $mc^2$ we immediately
deduce
$$m \approx 10^{-8}m_e$$
which agrees with (\ref{e12}) and (\ref{e13}). It also follows that $N \sim 10^{90}$,
which is correct. Moreover equating this energy of the quantum mechanical black hole
to $kT$, we get (cf. also (\ref{e12})
$$T \sim 1^o K,$$
which is the correct cosmic background temperature.\\
Alternatively, using (\ref{e12}) and (\ref{e13}) we get from (\ref{e14}), a
background radiation of a few millimeters wavelength, as required.\\
So we obtain not only the mass and the number of the neutrinos,
but also the correct cosmic background temperature, at one stroke.
\section{Discussion}
1) Hayakawa (cf.ref.\cite{r20}) in effect assumes the above neutrino mass and
equating the energy of oscillation of the background neutrino gas which is
in equilibrium under the Fermi gas pressure and gravitational attraction, with
the energy due to the weak Fermi interaction deduces the correct value of the
Fermi coupling constant $G_F$.\\
On the other hand, if the weak interaction is mediated by an intermediate
particle of mass $M$ and Compton wavelength $L$, we will get exactly as in the
model described in Section 2 for the electrons\cite{r21} (cf.ref.\cite{r8} also),
from the fluctuation of particle number $N$, on using (\ref{e13}),
\begin{equation}
g^2\sqrt{N} L^2 \approx mc^2 \sim 10^{-14},\label{e15}
\end{equation}
From (\ref{e15}), on using the value of $N$, we get,
$$g^2L^2 \sim 10^{-59}$$
This agrees with experiment and the theory of massless particles the neutrino
specifically acquiring mass due to interaction\cite{r22}, using the usual
value of $M \sim 100 Gev.$.\\
Alternatively, from (\ref{e15}), we get the correct value of $g^2/m^2_W$.\\
Thus a complete characterization of the weak interaction is possible.\\
2)The present value of the neutrino mass, as given by equation (\ref{e13}),
for example falls well short of the $10eV$ required to close the universe.
Thus there is no contradiction with the latest observations which indicate that
the universe is expanding for ever\cite{r23}.\\
3) Interestingly, if we use Bose-Einstein statistics, and equation (\ref{e10})
for solar neutrinos, rather than for the background neutrinos as we have done,
their number would need to be halved, as in the solar neutrino puzzle.\\
4) The preceding considerations do not contradict Hayakawa's earlier work,
as seen above. Indeed the fact that the background neutrino temperature equals the Fermi
temperature would also explain the Bosonization effect. One way to see this
is as follows:\\
For a collection of Fermions, we know that the Fermi energy is given by
\cite{r19},
\begin{equation}
\epsilon_F = p^2_F/2m = (\frac{\hbar^2}{2m}) (\frac{6\pi^2}{v})^{2/3}\label{e16}
\end{equation}
where $v^{1/3}$ is the interparticle distance. On the other hand, in a different
context, for phonons, the maximum frequency is given by, (cf.ref.\cite{r19}).
\begin{equation}
\omega_m = c(\frac{6\pi^2}{v})^{1/3}\label{e17}
\end{equation}
This occurs for the phononic wavelength $\lambda_m \approx$ inter-atomic
distance between the atoms, $v^{1/3}$, being, again, the mean distance between
the phonons. $'c'$ in (\ref{e17}) is the velocity of the wave, the velocity
of sound in this case. The wavelength $\lambda_m$ is given by,
$$\lambda_m = \frac{2\pi c}{\omega_m}$$
We can now define the momentum $p_m$ via the de Broglie relation,
$$\lambda_m = \frac{\hbar}{p_m},$$
which gives,
\begin{equation}
p_m = \frac{\hbar}{c}\omega_m,\label{e18}
\end{equation}
We can next get the maximum energy corresponding to the maximum frequency
$\omega_m$ given by (\ref{e17}), which as is known is,
\begin{equation}
\epsilon_m = \frac{p^2_m}{2m} = \frac{\hbar^2}{2m}(\frac{6\pi^2}{v})^{2/3}\label{e19}
\end{equation}
Comparing (\ref{e16}) and (\ref{e19}), we can see that $\epsilon_m$ and $p_m$
exactly correspond to $\epsilon_F$ and $p_F$.\\
The Fermi energy in (\ref{e16}) is obtained as is known by counting all single particle
energy levels below the Fermi energy $\epsilon_F$ using Fermi-Dirac
statistics, while the maximum energy in (\ref{e19}) is obtained by counting
all energy levels below the maximum value, but by using Bose-Einstein
statistics (cf.ref.\cite{r19}).\\
We can see why inspite of this, the same result is obtained in both cases.
In the case of the Fermi energy, all the lowest energy levels below
$\epsilon_F$ are occupied with the Fermionic occupation number $<n_p > = 1$,
$p < p_F.$ Then, the number of levels in a small volume about $p$ is $d^3p$.
This is exactly so for the Bosonic levels also. With the correspondence given
in (\ref{e18}), the number of states in both cases coincide and it is not
surprising that (\ref{e16}) and (\ref{e19}) are the same. In effect,
Fermions below the Fermi energy have a strong resemblance to phonons. This
is reminiscent of Fermi-Bose transmutation\cite{r24}.\\

\end{document}